\documentclass[11pt,twoside]{article}

\usepackage{asp2004}
\usepackage{epsf}
\usepackage{psfig}
\usepackage{lscape}

\begin{document}
\title{Differential rotation in early type stars}
\author{J. Zorec$^1$, Y. Fr\'emat$^2$, A. Domiciano de Souza$^3$} 
\affil{$^1$Institut d'Astrophysique de Paris, UMR7095 CNRS, Univ. P\&M Curie}
\affil{$^2$Royal Observatory of Belgium}
\affil{$^3$ Laboratoire Univ. d'Astrophysique de Nice (LUAN), UMR6525 CNRS}

\begin{abstract}
 Using 2D models of rotating stars, the interferometric measurements of 
$\alpha$~Eri and its fundamental parameters corrected for gravitational 
darkening effects we infer that the star might have a core rotating 2.7 times 
faster than the surface. We explore the consequences on spectral lines 
produced by surface differential rotation combined with the effects due to 
a kind of internal differential rotation with rotational energies higher than
allowed for rigid rotation which induce geometrical deformations that do not 
distinguish strongly from those carried by the rigid rotation. 
\end{abstract}
\vspace{-0.5cm}

\section{Models of internal differential rotation. The case of 
$\alpha$Eri}\label{model}  

\begin{figure}[]
\centerline{\psfig{file=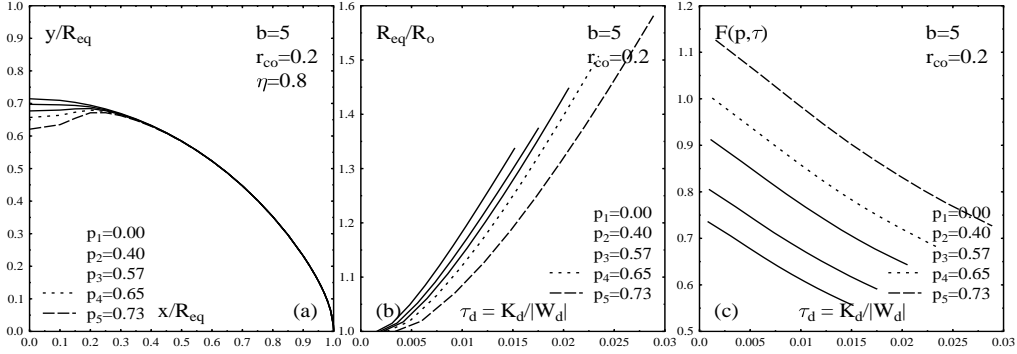,width=13.5truecm,height=5truecm}}
\caption[]{a) Meridian cuts of models with $b=5$, $r_{\rm co}=$ 0.2, $\eta=0.8$
and several values of $p$. b) Equatorial radius $R_{\rm eq}/R_o$ against 
$\tau_d=$ $K/|W|$ and $p$. c) Function $F(p,\tau_d)$ against $\tau_d$ and $p$}
\label{f1}
\end{figure}

 An initial rigid rotation in the ZAMS switches in some $10^4$ yr into an 
internal differential rotation (IDR) (Meynet \& Maeder 2000, MM2000). Nothing 
precludes then that IDR be present since the pre-main sequence phases implying
a total angular momentum higher than allowed for rigid rotation. Inspired by 
the internal rotation laws obtained in MM2000, we adopt the step-like 
rotational law: $\Omega(\varpi)\!=$ $\Omega_{\rm co}[1-p.e^{-a.\varpi^b}]$, 
where $\varpi\!=$ distance to the rotation axis; $\Omega_{\rm co}\!=$ core 
angular velocity; $p$ determines $\Omega_{\rm co}/\Omega_{\rm surf}$; 
$a\!=\!c(r_{\rm co})$ where $r_{\rm co}\!=\!R_{\rm core}/R_{\rm eq}\!=$ 
distance at which $\Omega\!=$ $(1/2)\Omega_{\rm co}$; $b$ determines the 
steepness of the drop from $\Omega_{\rm co}$ to $\Omega_{\rm surf}$. From 
MM2000 we adopt $b\!=\!5$, so that the rotational law is stable against 
axi-symmetric perturbations [$\partial{j}/\partial\varpi\!>\!0$; 
$j\!=\!\Omega(\varpi)\varpi^2$] for $p\!\la\!p_{\rm max}(b\!=\!5)\!=\!0.73$.
The geometrical deformation of stars is obtained by solving the 2D Poisson 
equation (Clement 1979):
\begin{equation}
\Delta\Phi_G(\theta,\varpi) = -4\pi\rho(\theta,\varpi)
\label{poisson}
\end{equation}
\noindent using barotropic relations $P=P[\rho(r)]$ of stellar interiors 
without rotation in different evolutionary stages (Zorec et al. 1989, Zorec 
1992). Fig.~\ref{f1}a shows stellar meridian cuts for $\eta=$ $\Omega^2_{\rm 
surf}R^3_{\rm eq}/GM$, several values of $p$ and $r_{\rm co}=0.2$. 
Fig.~\ref{f1}b shows $R_{\rm eq}/R_o$ ($R_o=$ stellar radius at rest) against
$\tau_d=$ ${\cal K}/|{\cal W}|$ (${\cal K}=$ rotational energy; ${\cal W}=$ 
gravitational potential energy).\par
 We considered that the apparent polar radius $R_{\rm po}^{\rm app}$ of 
$\alpha$ Eri is well determined (Domiciano de Souza et al. 2003, Vinicius et 
al. 2005), while the equatorial radius $R_{\rm eq}$ is obtained by asking the 
equivalent circular (determined spectrophotometrically and from visible 
interferometry) and the actual elliptical areas of the apparent stellar disc 
be equal. The relation between apparent and true stellar radii ratios: 
$R_{\rm po}^{\rm app}/R_{\rm eq}=$ $\{1-[1-(R_{\rm po}^{\rm true}/R_{\rm 
eq})^2]^2\sin^2i\}^{1\over2}$ leads to $F(p,\tau)$ that can be evaluated with
observed quantities and whose theoretical counterpart is (see Fig.~\ref{f1}c):
\begin {equation}
F(p,\tau) = \left[1-\left(R_{\rm po}/R_{\rm 
eq}\right)^2\right]/\left(V_{\rm e}/V_{\rm c,r}\right)^2
\label{funcf}
\end{equation}
\noindent where $V_{\rm eq}=$ equatorial velocity; $V_{\rm c}=$ critical 
velocity at rigid rotation. Thanks to interferometric data, we can look for 
models that reproduce the following observables: $R_{\rm eq}/R_{\odot}$, 
$R_{\rm eq}/R_{\rm po}$, $V\!\sin i$ corrected for gravity darkening effects
(Fr\'emat et al. 2005) and $F(p,\tau_d)$. Thus, we infer:
\begin{equation}
\left.\begin{array}{lclrclrcl}
   p&=& 0.624\pm0.001 &      \tau &=& 0.014\pm0.001 & i&=& 52^o\\
\eta&=& 0.69\pm0.07   & V_{\rm eq}&=& 308\pm16 {\rm km/s} & \Omega_{\rm 
co}/\Omega_{\rm surf}& = & 2.7 \\
\end{array}
\right\}
\end{equation}
\noindent For rigid rotation it is $p=0.0$ and $\tau_d\!\la\!0.015$.

\section{Combined effects of internal and surface differential rotations on
spectral lines}

\begin{figure}[]
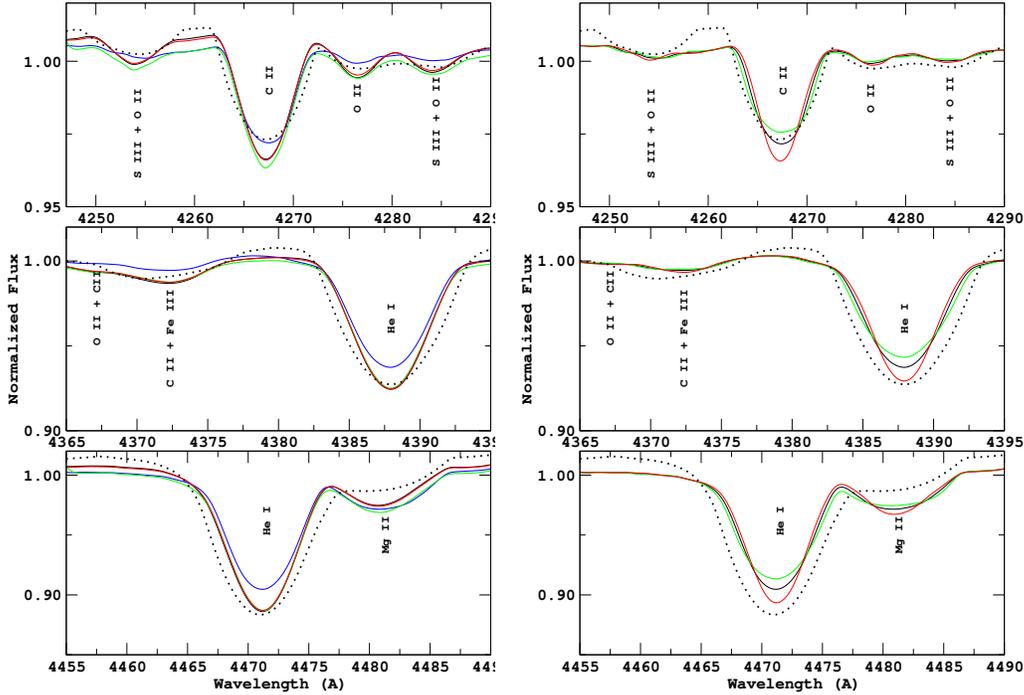

\centerline{\psfig{file=jzorec2_fig2a.eps,width=6.7truecm,height=9.2truecm}
\psfig{file=jzorec2_fig2b.eps,width=6.7truecm,height=9.2truecm}}
\caption[]{(left) Synthetic spectra produced by a star with a surface at rigid 
rotation, $V_{\rm surf}\!\sin i\!=\!$ 360 km/s, $\eta\!=\!$ 0.95, $p\!=\!0.7$
[rest $T_{\rm eff}\!=\!$ 20000 K, $\log\!g\!=\!3.5$] ($dots\!=$ spherical and
no GD; $bold\ black\!=$ rigid rotator with GD; $thin\ black\!=$ IDR with 
$r_{\rm co}\!=\!0.3$; $grey\!=$ IDR with $r_{\rm co}\!=\!0.2$. (right) Spectra 
from the same star as in the left but including surface differential rotation:
$dots\!=$ no GD and $\alpha\!=\!0.0$; $bold\ black\!=$ GD+$\alpha\!=\!0.0$; 
$thin\ black\!=$ GD+$\alpha\!=\!0.5$; $grey\!=$ GD+$\alpha\!=\!-0.5$)}
\label{f2}
\end{figure}

 The emitted spectrum of a differential rotator depends sensitively on the
value of $R_{\rm core}/R_{\rm eq}$. Fig.~\ref{f2}(left) shows spectral 
lines in the $\lambda\lambda4250-4490$~\AA\ interval produced by a star
[rest $T_{\rm eff}\!=\!$ 20000 K, $\log\!g\!=\!$ 3.5] with a surface at rigid 
rotation, $V_{\rm surf}\!\sin i=$ 360 km~s$^{-1}$, $\eta=$ 0.95, $p=0.7$ 
($dots\!=$ spherical star without gravitational darkening (GD); 
$bold\ black\!=$ rigid rotator with GD; $thin\ black\!=$ differential rotator
with $r_{\rm co}\!=\!0.3$; $grey\!=$ differential rotator with $r_{\rm 
co}\!=\!0.2$. If we consider that the angular velocity against the co-latitude 
$\theta$ is:
\begin{equation}
\Omega_{\rm surf}(\theta) = \Omega_{\rm surf,eq}(1-\alpha\cos^2\theta)
\end{equation}
\noindent and use the same parameters as in Fig.~\ref{f2}(left) but 
$\alpha\neq0$ we obtain the line profiles shown in Fig.~\ref{f2}(right) 
($dots\!=$ no GD and $\alpha=0.0$; $bold\ black\!=$ GD+$\alpha=0.0$; 
$thin\ black\!=$ GD+$\alpha=0.5$; $grey\!=$ GD+$\alpha=-0.5$). We can show that
for $R_{\rm core}/R_{\rm eq}\to0$ an apparent rigid rotation regime is 
recovered, but for a higher $\tau_d$ ratio. According to $\tau_d$ a given line 
equivalent width may imply different stellar fundamental parameters. The 
sensitivity to the stellar deformation and to the related GD depends on the 
spectral line and the inclination $i$. Due to a higher change of the polar 
radius and consequent higher local effective temperatures than for a rigid 
rotation, lines can be deepened, shallowed, or self-reversed. The differences
in the equivalent widths carry uncertainties on the chemical abundance 
determinations. The surface differential rotation in gravity darkened stars 
carry deepening of line profiles if $\alpha\!<\!0$ and they are shallowed if 
$\alpha\!>\!0$. Other related subjects can be found in 
http://www2.iap.fr/users/zorec/.
 
\end{document}